\documentclass{elsart3-1}

\usepackage{amssymb}

\usepackage[english,francais]{babel}
\usepackage[T1]{fontenc}
\usepackage[latin1]{inputenc}
\usepackage{graphicx}

\newtheorem{e-proposition}[theorem]{Proposition}

\newtheorem{e-definition}[theorem]{Definition\rm}

\renewcommand{\theequation}{\arabic{equation}}
\def\og{\leavevmode\raise.3ex\hbox{$\scriptscriptstyle\langle\!\langle$~}}
\def\fg{\leavevmode\raise.3ex\hbox{~$\!\scriptscriptstyle\,\rangle\!\rangle$}}

\begin{document}

\begin{frontmatter}

% Titre, auteurs et adresses

% utiliser la commande \thanksref dans \title, \author ou \address
%     pour les notes en bas de page ;

% utiliser la commande \ead pour l'adresse e-mail de chaque auteur
%    (aprËs la commande \auteur) ;

% \author{Name\thanksref{label2}}
\selectlanguage{francais}
\title{Ergodicité, collage et transport anomal}

% utiliser les Ètiquettes pour indiquer l'adresse de chaque auteur,
%     s'il y a plusieurs adresses


\author[authorlabel1]{Xavier Leoncini}
\ead{Xavier.Leoncini@cpt.univ-mrs.fr}
\author[authorlabel1]{Cristel Chandre}
\ead{Cristel.Chandre@cpt.univ-mrs.fr}
\author[authorlabel2]{Ouerdia Ourrad}
\ead{omeziani@yahoo.fr}
\address[authorlabel1]{Centre de Physique Théorique, UMR 6207, Aix-Marseille Universités, Luminy, Case 907
F-13288 Marseilles cedex 9, France}
\address[authorlabel2]{Laboratoire de Physique Théorique, Université A. Mira - Targa Ouzemour
06000, Bejaia, Algerie}
% etc, etc

% Vous pouvez ajouter a la prochain ligne les dates
%   de reception, et d'acceptation, et apres le nom du presentateur

\medskip
\selectlanguage{francais}
\begin{center}
{\small Re\c{c}u le *****~; accept\'e apr\`es r\'evision le +++++\\
Pr\'esent\'e par £££££}
\end{center}

\begin{abstract}
\selectlanguage{francais}
% Texte du rÈsumÈ en franÁais
Nous nous intéressons à la convergence vers sa moyenne spatiale ergodique de la moyenne
temporelle d'une observable d'un flow hamiltonien  à un degré et demi de liberté avec
espace des phases mixte. L'analyse est faite au travers de l'évolution de la distribution des
moyennes en temps fini d'un ensemble de conditions initiales sur la même composante ergodique.
 Un exposant caractérisant la vitesse de convergence est défini. Les résultats
indiquent que pour le système considéré la convergence évolue en $t^{\alpha}$,
avec $\alpha=0.45$ pour  alors qu'elle
évolue en $t^{1/2}$ lorsque la dynamique est globalement chaotique dans l'espace des phases. De
même une loi $\alpha=1-\beta/2$ reliant cet exposant
 $\alpha$ à l'exposant caractéristique du deuxième
moment associé aux propriétés de transport  $\beta$
est proposée et est vérifiée pour les cas considérés.
{\it Pour citer cet article~: A. Nom1, A. Nom2, C. R.
Mecanique 333 (2005).}

\vskip 0.5\baselineskip

\selectlanguage{english}
% Texte du rÈsumÈ en franÁais
\noindent{\bf Abstract}
\vskip 0.5\baselineskip
\noindent
{\bf Ergodicity, stickiness and anomalous transport }
We consider the problem of convergence towards  spatial ergodic average of the  time
 average of an observable defined for a one and half degree of freedom Hamiltonian flow with mixed phase space.
The analysis is performed by analysing the evolution of the distribution of finite-time averages. 
An exponent characterising the ``speed of convergence'' is defined. Results indicate that for the
considered mixed case, the rate of convergence goes as $t^{\alpha}$,
with $\alpha=0.45$ while it goes as $t^{1/2}$ when the full phase space is chaotic.
 Moreover a formula linking  this characteristic exponent to the one corresponding
to transport properties $\beta$ is proposed $\alpha=1-\beta/2$ and  good agreement is found for the considered cases.
{\it To cite this article: A. Nom1, A. Nom2, C. R.
Mecanique 333 (2005).}

%Maintenant keywords/mots-clÈs, le premier doit venir de la liste publiÈe dans CR Mecanique
\vskip 0.5\baselineskip
\noindent{\small{\it Mots-cl\'es~:} Systèmes Dynamiques; Chaos Hamiltonien; Transport Anomal}}
\end{abstract}
\end{frontmatter}

% Maintenant la version abrÈgÈe en anglais, si prÈsente
\selectlanguage{english}
\section*{Abridged English version}
% Texte de la version abrÈgÈe en anglais
In this note we consider the problem of convergence to the ergodic average of an observable of a one and a half degree of freedom Hamiltonian flow, and its connection to transport properties. The considered observable is the norm of the speed $(\dot{p},\:\dot{q})$ defined  in phase space, and the transport properties of the associated variable, namely the length in phase space of trajectories.

Assuming ergodicity we conclude that the infinite time limit distribution of time averaged speeds (\ref{eq:speed}), is a Dirac.
Then when supposing that  we have a very chaotic system of Anosov type,  we can expect  from the central limit theorem that the finite time average distribution has a maximum peak which grows to infinity as $t^{1/2}$. We consider this last property of the height of the maximum of the probability distribution function of finite-time averages for a system with mixed phase space and use it to define a characteristic exponent $\alpha$, expecting an algebraic growth as $t^{\alpha }$.

In order to link this exponent to transport properties we make the crude approximation of neglecting the  tails (described in Fig~.\ref{fig:naif}) to  link  the exponent $\beta$ characterising the second moment of transport properties with $\alpha$ and obtain the law $\alpha=1-\beta/2$, which should be valid for Gaussian transport.

We then test the ideas numerically with the simple pertubed pendulum (\ref{eq:Hamilton}) for two different values of the parameters, one giving us more or less uniform chaos in a given region of phase space, the other one presenting a chaotic sea within which  regular islands are present (see Fig.~\ref{fig:Section-de-Poincar=E9}). Numerical results are made using the fifth order optimal symplectic scheme described in \cite{McLachlan92}, using $1024$ different initial conditions. Trajectories are computed for $10^6$ periods with a time step $\delta t=T/200$. Histograms have been computed with a resolution of $5000$ bins. The evolution of the averaged speed distribution  and the evolution of the maximum in log-log plot are represented in Fig.~\ref{fig:Densit=E9-de-probabilit=E9s}. One typically see that for both cases, the exponent $\alpha$ can be determined. We find $\alpha=1/2$ for the chaotic case while $\alpha=0.43$ for the mixed case.
More over we compute the transport properties for both system and compare the measured $\beta$'s with the ones expected from Eq.~(\ref{eq:lien_alpha-beta}). Results are presented in table~\ref{tab:R=E9capulatif-des-exposants}, and a good agreement is found with the extrapolated exponents  form the $\alpha=1-\beta/2$ rule.

\selectlanguage{francais}
% texte principale
\section{Introduction}
\label{}

Dans cette note nous considérons un flot hamiltonien  à un degré
et demi de liberté $H(p,q,t)$ et périodique dans le temps et nous nous
intéressons à la convergence de la moyenne temporelle d'une observable,
la norme de la vitesse dans l'espace des phases. Plus précisément
nous considérons l'évolution de la distribution des moyennes temporelles
d'un ensemble de  différentes conditions initiales
et étudions les propriétés d'ergodicité du système\cite{Collet2005}. Pour une condition
initiale donnée $i$, la vitesse moyenne au bout de $n$ périodes
s'écrit:\begin{equation}
\bar{v_{i}}(n)=\frac{1}{nT}\int_{0}^{nT}\sqrt{\dot{q}_{i}^{2}+\dot{p}_{i}^{2}}dt\:.\label{eq:speed}\end{equation}
On considère alors un ensemble de conditions initiales que l'on suppose
équivalentes. Ainsi si la dynamique est ergodique chaque $\bar{v_{i}}(n)$
va tendre vers la moyenne spatiale quand $n\rightarrow\infty$, et
donc la distribution des $\bar{v_{i}}(n)$ pour l'ensemble des conditions
initiales équivalentes tendra vers une distribution de Dirac. L'objet
de ce travail est d'étudier comment cette distribution est approchée
(si elle existe) et de définir un exposant caractéristique. Nous considérons
dans un premier temps un cas {}``idéal'' d'un système Hamiltonien
globalement chaotique pour lequel les résultats observés sont conformes
au théorème de la limite centrale, pour considérer ensuite un système
où l'espace des phases est mixte, c'est-à-dire un système pour lequel
nous avons une mer chaotique peuplée d'îlots de stabilité au sein
desquels la dynamique est régulière. Pour ce dernier système nous
considérons uniquement des conditions initiales dans la mer chaotique.

\section{Définitions}

Nous nous proposons de définir un exposant caractéristique à partir
des considérations suivantes. Tout d'abord imaginons  un système
 chaotique, de type Anosov, tel que l'on puisse appliquer le théorème
de la limite centrale. Dans ce cas la variable $y_{i}(n)=\sqrt{n}\bar{v_{i}}(n)$
pour $n$ grand va se comporter comme une variable aléatoire distribuée
sur une gaussienne d'écart type fini. Savoir comment la distribution
converge vers la distribution limite est l'objet de la théorie des
grandes déviations mais cette propriété de convergence ergodique implique
que la distribution des $\bar{v_{i}}(n)$ va tendre vers un Dirac
(une gaussienne d'écart type nul). De même on peut anticiper que l'écart
type de cette distribution va décroître en $1/\sqrt{n}$ et que par
conservation des aires (la probabilité totale est conservée) le maximum
de la distribution $\rho_{max}(n)$ va croître en $\sqrt{n}$\cite{Chazottes99}. %
\begin{figure}
\begin{centering}
\includegraphics[width=8cm]{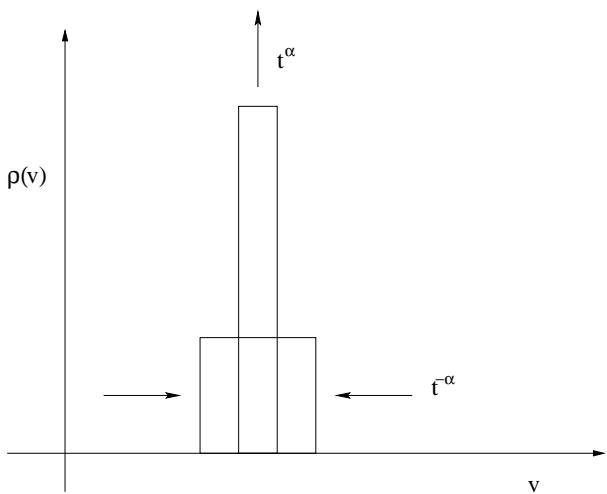}
\par\end{centering}

\caption{Estimation de la relation entre l'écart type et le maximum d'une distribution\label{fig:naif}}
\end{figure}
Cette dernière propriété permet de définir un exposant \begin{equation}
\rho_{max}(n)\sim n^{\alpha}\label{eq:definition_alpha}\end{equation}
 (avec $\alpha=1/2$), que nous pouvons à priori généraliser pour
des systèmes moins chaotiques avec la possibilité d'avoir $\alpha\ne1/2$.

Intéressons nous maintenant aux propriétés de transport du système
de type Anosov, en mesurant la longueur des trajectoires dans l'espace
des phases, et donc à l'évolution de la distribution des variables
$s_{i}(n)=\sqrt{n}Ty_{i}(n)$. Par les mêmes considérations que celles
faites précédemment nous attendons un comportement diffusif avec une
croissance linéaire de la variance. Maintenant considérons un système
où le transport est anomal comme par exemple un système hamiltonien
à un degré et demi de liberté \cite{LKZ01}. Le transport est dit
anomal, dans le sens où la variance $M_2=\langle(s_{i}(n)-\langle s_{i}(n)\rangle)^{2}\rangle\sim n^{\beta}$
a un comportement non-linéaire. Cette anomalie fait suite aux effets
de mémoires engendrés par le phénomène de collage autour d'îlots
de stabilité \cite{Zaslavsky2002,LKZ01} qui engendrent une décroissance
en loi de puissance des queues de la distribution de $s_{i}(n)$.
Dans ce cas en faisant le même type de raisonnement que celui illustré
sur la figure~\ref{fig:naif},  un raisonnement négligeant la
contribution des queues de distribution et en utilisant la proportionalité
$\bar{v_{i}}(n)$=$s_{i}(n)/nT$, nous obtenons \begin{equation}
\alpha=1-\beta/2\:.\label{eq:lien_alpha-beta}\end{equation}
 Ainsi on qualifiera (par abus de langage) le transport de diffusif
si $\alpha=1/2$, mais aussi on pourra anticiper un comportement anomal avec
$\alpha<1/2$ si le transport est superdiffusif et $\alpha>1/2$ si
il est sous-diffusif. De même la condition de convergence ergodique
$\alpha\ge0$ implique dans le cadre où la loi (\ref{eq:lien_alpha-beta})
s'applique $\beta\le2$.

\section{Résultats}

Afin de tester les idées et définitions énoncées auparavant nous considérons
le système hamiltonien  du pendule perturbé, qui est également \emph{}le
hamiltonien décrivant l'évolution d'une particule chargée évoluant
dans le potentiel créé par deux ondes {}``électrostatiques'' dans
le référentiel lié à l'une d'elle \cite{Zaslavsky68,Escande85,Chandre-Jauslin2002}:

\begin{equation}
H=\frac{p^{2}}{2m}+A\left(\cos(k_{1}q)+\varepsilon\cos(k_{2}q-\omega t+\varphi)\right)\:,\label{eq:Hamilton}\end{equation}
où le couple $(p,q)$ désigne les variables conjuguées du hamiltonien.
Dans la suite et sans perte de généralité nous prendrons $m=A=k_{1}=1$,$\varphi=0$ 
et notons $k_{2}=k$ et considérerons les deux cas (a) $\nu=0.5$,
$k=1$, $\varepsilon=1$ et (b) $\nu=5$, $k=2$, $\varepsilon=1$.
\begin{figure}
\begin{centering}
\includegraphics[width=7.5cm]{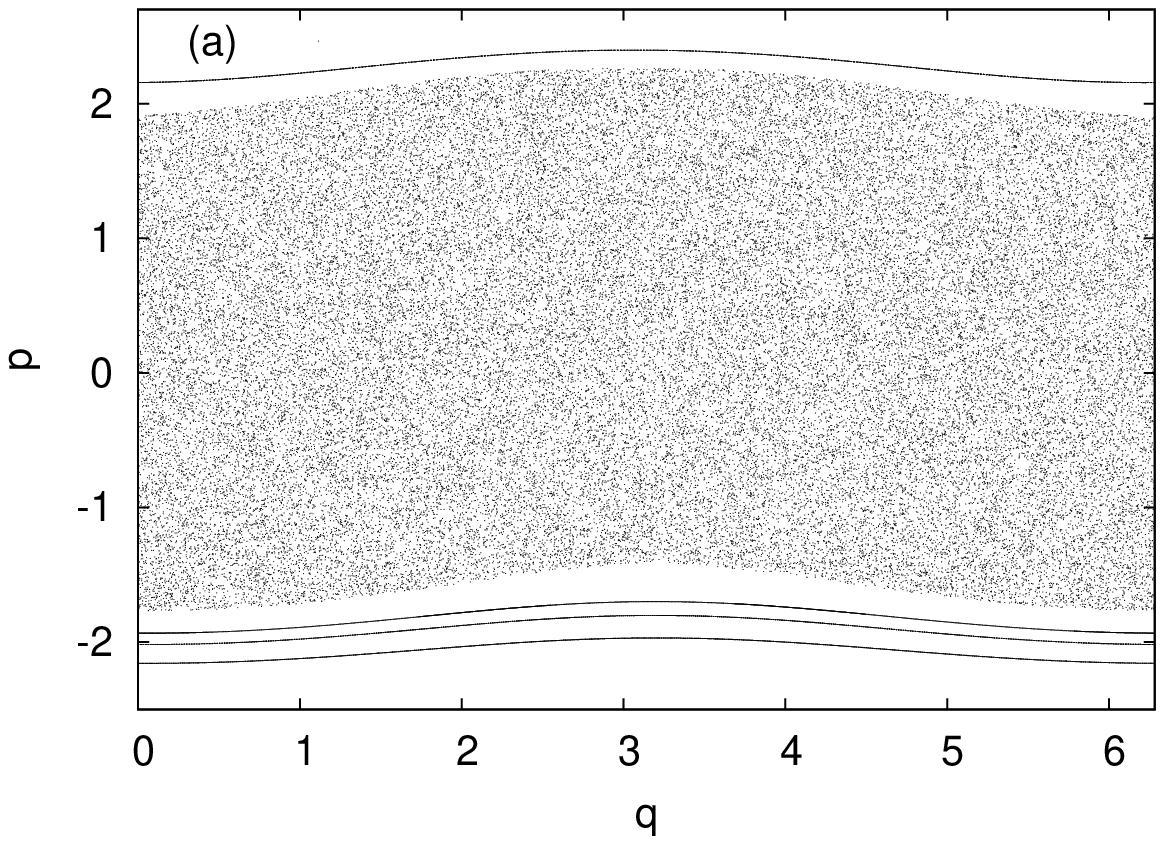}\includegraphics[width=7.5cm]{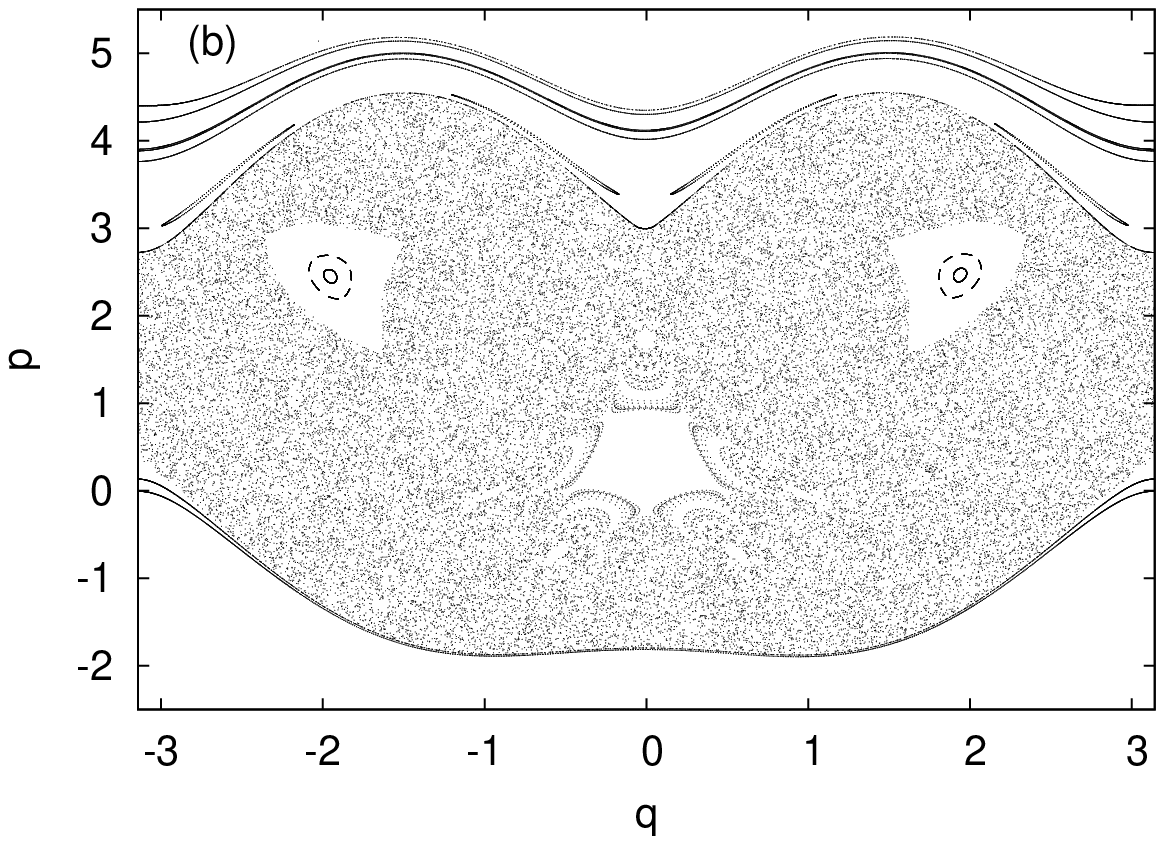}
\par\end{centering}

\caption{Section de Poincaré du hamiltonien (\ref{eq:Hamilton}) pour les
paramètres suivants: (a) $\nu=0.5$, $k=1$, $\varepsilon=1$ et (b)
$\nu=5$, $k=2$, $\varepsilon=1$.\label{fig:Section-de-Poincar=E9}}
\end{figure}
 Nous pouvons noter que sur les différentes sections de Poincaré représentées
en Fig.~\ref{fig:Section-de-Poincar=E9} que le cas (a) correspond
à une zone chaotique étendue délimitée par des tores de rotation réguliers de type KAM, tandis
que pour le cas (b) la zone chaotique est peuplée d'îlots au sein
desquels des trajectoires régulières sont présentes.

%
%\begin{figure}
%\begin{centering}
%\includegraphics[width=7.5cm]{Fig3a}\includegraphics[width=7.5cm]{Fig3b}
%\par\end{centering}
\begin{figure}
\begin{centering}
\includegraphics[width=7cm]{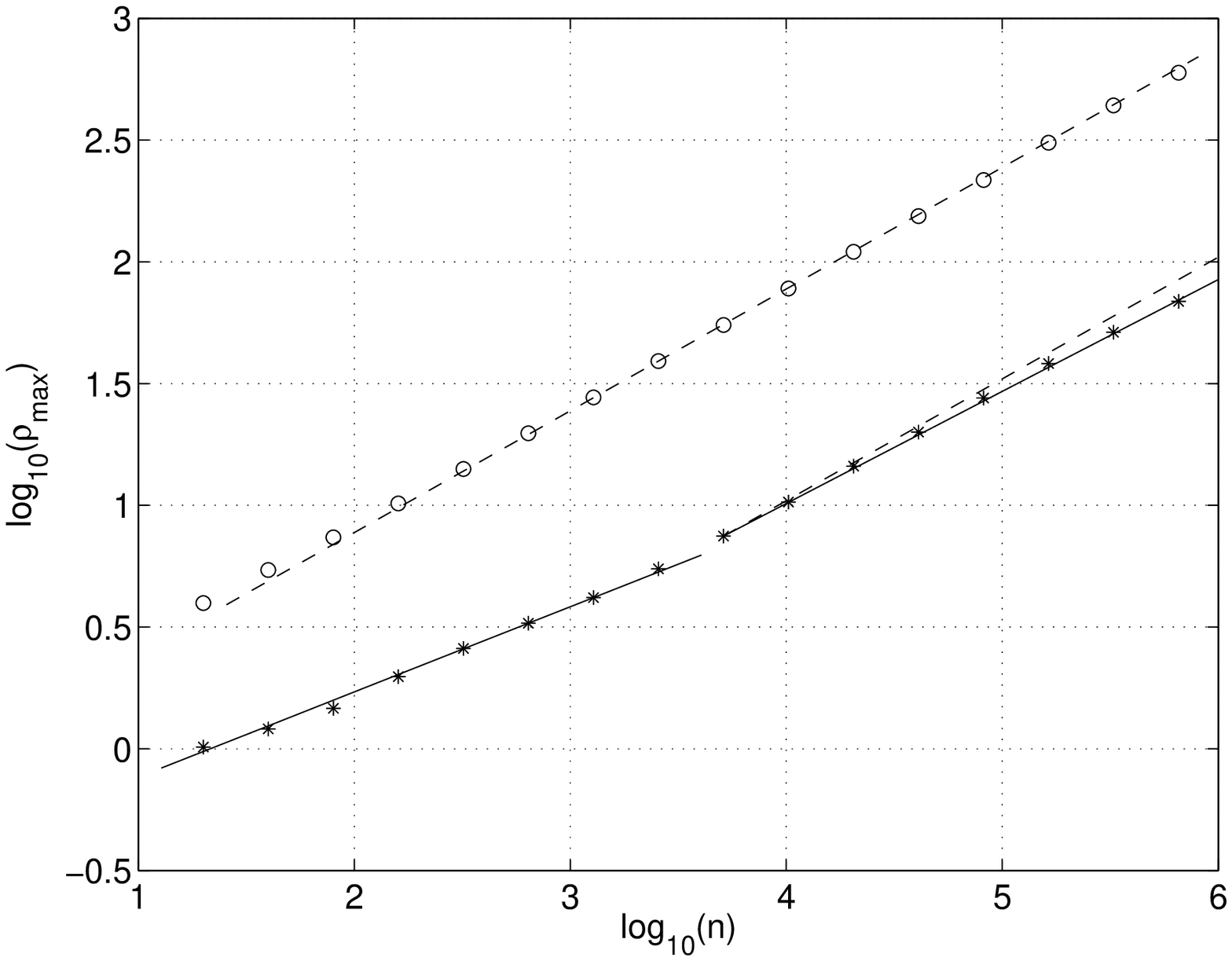}\\
\includegraphics[width=3.5cm]{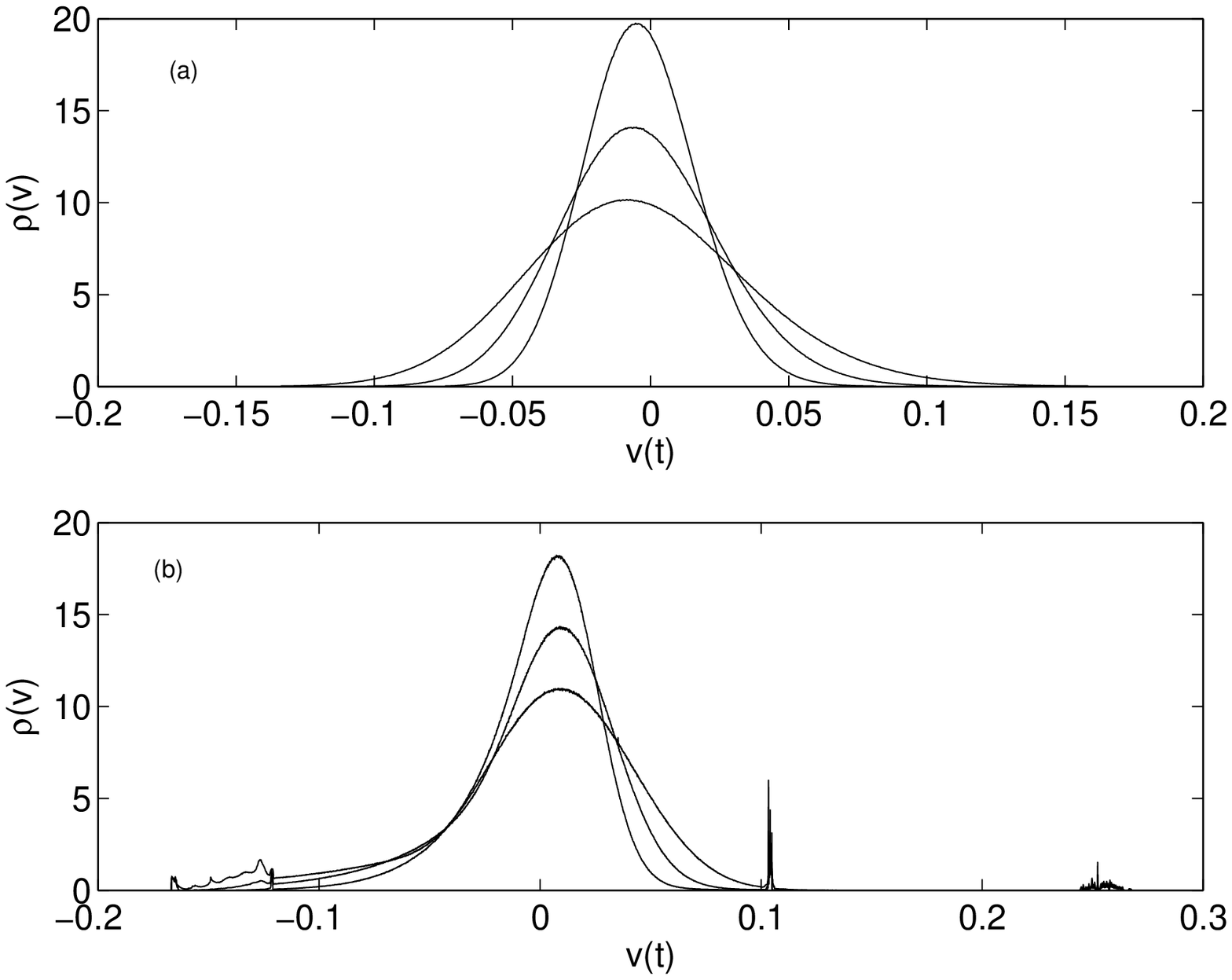}\includegraphics[width=3.5cm]{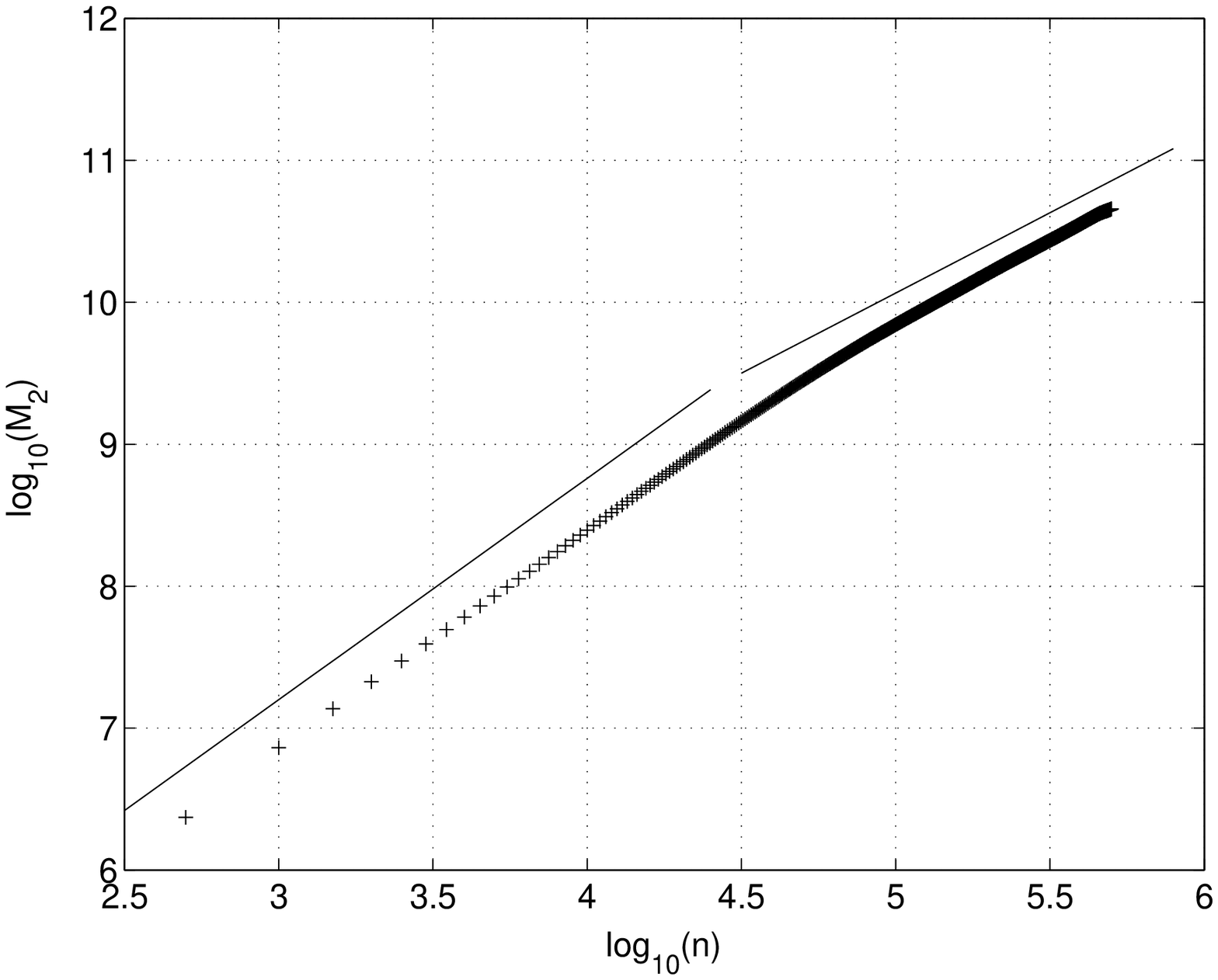}
\par\end{centering}

\caption{Haut: Évolution du maximum de la distribution des vitesses moyennes
$\rho_{max}(n)$ en fonction du nombre de périodes pour les deux cas
considérés. L'exposant $\alpha$ caractérisant la {}``convergence
vers l'ergodicité'' $\rho_{max}(n)\sim n^{\alpha}$ est mesuré à
$\alpha\approx0.5^{-}$ pour le cas (a) et pour le cas (b) on note
deux pentes $\alpha\approx0.35$ pour $n<2.5\ 10^{3}$ et $\alpha\approx0.45$
pour $2.5\ 10^{3}<n<6.5\ 10^{5}$, les lignes en pointillé ont une pente 1/2.
Bas: A gauche, densité de probabilité des vitesses moyennes pour les cas
(a) et (b) pour différents temps ($\tau=160,\;320,$ et $640$ périodes).
Les distributions ont été calculées avec une résolution de $5000$
points à partir de $1024$ trajectoires calculées pendant $10^{6}$
périodes. Le pas de temps est $\delta t=T/200$. A droite, évolution du second moment de la distribution des
longueurs parcourues $M_2$, les pentes mesurées sont $\beta=1.55$ puis  $\beta=1.13$ aux temps longs.
 \label{fig:Densit=E9-de-probabilit=E9s}}
\end{figure}
Nous suivons alors l'évolution de la distribution des $\bar{v_{i}}(n)$,
pour cela nous considérons un ensemble de conditions initiales prises
dans la mer chaotique. L'évolution vers un Dirac de cette distribution
pour les cas (a) et (b) et l'évolution du maximum de la distribution
sont représentées sur la figure~\ref{fig:Densit=E9-de-probabilit=E9s}.
A la différence du cas (a) nous notons la présence de pics secondaires
dans la distribution pour le cas (b). Ces pics s'expliquent par le
phénomène de collage autour des îlots \cite{LKZ01}. Ce phénomène n'est quasiment pas observé pour presque toutes les conditions initiales dans le cas (a) (une trajectoire se met à coller sur le tore supérieur après $10^5$ périodes).
 Numériquement, nous avons intégré
la dynamique en utilisant le schéma symplectique d'ordre cinq optimal décrit dans \cite{McLachlan92}.
Nous avons considéré un ensemble de $1024$ trajectoires différentes que nous avons calculées pendant
$10^6$ périodes avec un pas de temps $\delta t=T/200$.

Nous pouvons noter  vu les résultats présentés sur la  Fig.~\ref{fig:Densit=E9-de-probabilit=E9s}
que la caractérisation de l'exposant $\alpha$ est relativement aisée,
les lois étant assez linéaires sur plusieurs ordres de grandeurs.
\begin{table}
\begin{centering}
\begin{tabular}{|c|c|c|}
\hline 
&
(a)&
(b)\tabularnewline
\hline 
$\alpha$&
$0.5$&
$0.45$\tabularnewline
\hline 
$\beta$&
$1$&
$1.13$\tabularnewline
\hline 
$1-\beta/2$&
$0.5$&
$0.44$\tabularnewline
\hline
\end{tabular}
\par\end{centering}

\caption{Récapitulatif des exposants observés.\label{tab:R=E9capulatif-des-exposants}}
\end{table}
 De même la mesure de l'exposant caractéristique du transport associé
à $s_{i}$, calculé pour les presque toutes $1023$ trajectoires pour le cas (a) et toutes les trajectoires dans le cas (b) (voir tableau~\ref{tab:R=E9capulatif-des-exposants})
montre que la loi approximative reliant les deux exposants est bien
vérifiée pour les temps long, alors qu'elle ne l'est pas pour des temps courts dans le cas ou l'espace des phases est mixte.

Dans cette note nous avons défini un exposant caractéristique lié à la convergence vers la moyenne ergodique d'une observable.
Pour les cas considérés, la connaissance de cet exposant permet de prédire la nature du transport et ce même pour un cas où l'espace des phases est mixte. Du point de vue du transport anomal nous pensons que cette approche est complémentaire de celle liée à la notion de jets chaotiques \cite{LZ03} et de $\varepsilon-$complexité \cite{Afraimovich03}.

\section*{Remerciements}
X. L. tient à remercier Bastien Fernandez et Ricardo Lima, pour leurs nombreuses
remarques, conseils et corrections lors de l'élaboration de ce manuscrit.~\\
Ce travail s'inscrit au sein du contrat LRC DSM-06-35.

%\bibliographystyle{unsrt}
%\bibliography{/home/leoncini/texte/articles/bibtex/transport}

% etc, etcAfraimovich03

% Les remerciements sont dans une section, sans numÈrotation

%\section*{Remerciements}
% Remerciements - texte ici

%\begin{thebibliography}{00}
% essayez a utiliser le systeme 'bibitem',
%   avec les references en ordre de citation dans le texte
% Texte
%\bibitem{label2}
%
%\bibitem{}

%\end{thebibliography}

\end{document}